\documentclass[fleqn,usenatbib]{mnras}
\usepackage[T1]{fontenc}
\usepackage{ae,aecompl}
\usepackage{graphicx}   
\usepackage{amsmath}    
\usepackage{amssymb}    

\usepackage[dvipsnames]{xcolor}

\title{Hot X-ray Onsets of Solar Flares}

\author[H.S. Hudson et al.]{
\begin{tabular}{lll}
Hugh S. Hudson,$^{1,2}$\thanks{hugh.hudson@glasgow.ac.uk (HSH)}
Paulo J. A. Sim\~oes,$^{3,1}$
Lyndsay Fletcher,$^{1,4}$\\
Laura A. Hayes,$^{5}$
Iain G. Hannah$^{1}$
\end{tabular}
\bigskip
\\
$^{1}$SUPA School of Physics and Astronomy, University of Glasgow, Glasgow G12 8QQ, UK\\
$^{2}$Space Sciences Laboratory, U.C. Berkeley, CA USA\\
$^{3}$Centro de R\'adio Astronomia e Astrof\'isica Mackenzie,  Escola de Engenharia, Universidade Presbiteriana Mackenzie, S\~ao Paulo, Brazil\\
$^{4}$Rosseland Centre for Solar Physics,  University of Oslo, P.O.Box 1029 Blindern, NO-0315 Oslo, Norway\\
$^{5}$Solar Physics Laboratory, Code 671, Heliophysics Science Division, NASA Goddard Space Flight Center, Greenbelt, MD 20771, USA
}

\date{June 23, 2020}

\begin{document}

\maketitle

\begin{abstract}
The study of the localized plasma conditions before the impulsive phase of a solar flare can help us understand the physical processes that occur leading up to the main flare energy release. 
Here, we present evidence of a hot X-ray `onset’ interval of enhanced isothermal plasma temperatures in the range of 10-15~MK up to tens of seconds prior to the flare’s impulsive phase. 
This `hot onset’ interval occurs during the initial soft X-ray increase and prior to the detectable hard X-ray emission. 
The isothermal temperatures, estimated by the Geostationary Operational Environmental Satellite (GOES) X-ray sensor, and confirmed with data from the Reuven Ramaty High Energy Solar Spectroscopic Imager (RHESSI), show no signs of gradual increase, and the `hot onset' phenomenon occurs regardless of flare classification or configuration. 
In a small sample of four representative flare events we identify this early hot onset soft X-ray emission mainly within footpoint and low-lying loops, rather than with coronal structures, based on images from the Atmospheric Imaging Assembly (AIA).
We confirm this via limb occultation of a flaring region. 
These hot X-ray onsets appear before there is evidence of collisional heating by non-thermal electrons, and hence they challenge the standard flare heating modeling techniques.
\end{abstract}

\section{Introduction}\label{sec:intro}

Often flare initiation, as seen in soft X-ray (SXR) data from the X-ray Sensor (XRS) on \textit{Geostationary Operational Environmental Satellite} (GOES), begins with a slow `precursor' development phase.
This can sometimes be identified with non-thermal activity \citep[e.g.][]{2003A&A...399.1159F} or with
non-thermal velocity distributions \citep{2013ApJ...774..122H}.
The preflare interval is often also taken as evidence for a `pre-heating' phase in which a gradual process heats a volume
of flare plasma without a detectable hard X-ray signature \citep[e.g.][]{1985ApJ...298..887C}, implying a very low flux of non-thermal electrons, if any.
Systematic studies of soft X-ray images suggested that in most cases any precursor source could not be directly identified with
the main flare \citep{1998SoPh..183..339F,2008ASPC..397..130H}, appearing near but not exactly at the flare site.


Previous conclusions about the relationship between `precursors' and flares mostly have dealt with image structure.
Here we study the X-ray spectral evolution, focusing on a sample of four representative events.
We examine flare \textit{onset} emission, where we define the term `onset' as the pre-flare interval during which elevated GOES soft X-ray flux is detected, but prior to the detection of any elevated hard X-ray (HXR) emission (at $>$25~keV for stronger events, and 12-25~keV for weaker ones) by the Reuven Ramaty High Energy Solar Spectroscopic Imager \citep[RHESSI, ][]{2002SoPh..210....3L}. 
Our main finding is that the GOES isothermal temperatures  are significantly elevated from the very beginning of the onset phase, i.e. well before we have evidence for collisional heating by non-thermal electrons.
We  cannot preclude the possibility of undetectable HXR emission, especially with a softer spectrum, in the onset time interval.
We note that \cite{2011ASInC...2..297A} had already reported similar phenomena via the independent dataset from the \textit{Solar X-Ray Spectrometer} (SOXS) spectrometer experiment \citep{2006JApA...27..175J}, in a sample of 13 events.

We have used extreme ultraviolet (EUV) images to search out the spatial patterns of the onset sources (Section~\ref{sec:images}). 
We have also studied the RHESSI data for the four sample events (Section~\ref{sec:rhessi}), finding satisfactory qualitative agreement during the flare development, specifically in matching the isothermal-fit GOES parameters with the more complete spectroscopy possible with RHESSI. 

\section{Data}

\subsection{GOES soft X-ray data}

The GOES series of missions has provided soft X-ray measurements via its X-ray Sensor (XRS) instrument in two nominal wavelength bands, (1--8~\AA\ and 0.5-4~\AA), for many decades now. 
Such observations began as early as 1960 with ionization chambers on board SOLRAD and other satellites \citep{1974JATP...36..989D,1985SoPh...95..323T,2005SoPh..227..231W}. 
The two passbands of the GOES/XRS, 1--8~and 0.5--4~\AA, allow for the determination of an isothermal temperature and emission measure, interpreted here in terms of the CHIANTI atomic-physics package \citep{1997A&AS..125..149D} as implemented in the SolarSoft \citep{1998SoPh..182..497F} code GOES\_TEM.pro.
These parameters usually describe the coronal part of the flare, and specifically the plasma trapped in the system of magnetic loops made visible in soft X-rays by the injection of new plasma expanding upwards from the lower atmosphere due to the sudden energy release. The GOES data also have sufficient sensitivity and signal contrast to study the onset phase of a flare, often many minutes prior to the impulsive phase \citep{1970ApJ...162.1003K}. 
In this paper we use these simple GOES/XRS observations to characterize the onset temperatures, at the earliest possible times permitted by the observations, and then follow up with EUV images, taken by the \textit{Atmospheric Imaging Assembly} \citep[AIA, ][]{LemenTitleAkin:2012} on board the \textit{Solar Dynamics Observatory} \citep[SDO, ][]{PesnellThompsonChamberlin:2012}.
We also examine a flare series in which limb occultation distinguishes the coronal and chromospheric components \citep[e.g.][]{1978ApJ...224..235H}.

Because the GOES data integrate over the whole disk (Sun-as-a-star), all of the concurrent soft X-ray sources will contribute to the background level for a given flare.
In principle there is no exact way to estimate this background level for such Sun-as-a-star observations, since an independent source(s) could occur at any time, in any active region that might be coincidentally present.
In the present work we estimate the flare background level by simply taking the local minimum of the 0.5-4~\AA\ channel immediately prior to the flare onset; the nearer the better. 
The actual epoch of the hot onset will depend upon flare brightness and detection threshold; for an X-class flare occurring in low-background conditions, the GOES photometers can already detect the source at a level 0.1-1\% of flare maximum flux.

We can also check the background source locations via the EUV images from SDO/AIA. Among its passbands, \cite{2013ApJ...771..104F} showed that the timeseries of the 131~\AA\ data provides the closest match to the GOES~1-8~\AA\ timeseries at least for the flare SOL2010-08-07T18:24 (M1.0). 
During flares, the AIA 131 \AA\ passband is dominated by the Fe {\sc xxi} 128.8 \AA\ and Fe {\sc xxiii} 132.9 \AA\ lines, formed at $\log T$ = 7.05 and 7.15, respectively, while the 94 \AA\ passband captures the emission from Fe {\sc xviii} 93.9 \AA\ and Fe {\sc xx} 93.8 \AA\, formed at $\log T$ = 6.85 and 7.0, respectively \citep[e.g.][]{ODwyerDel-ZannaMason:2010}.

\subsection{Event sample}

\begin{figure*}
\centering
\includegraphics[width=0.9\textwidth]{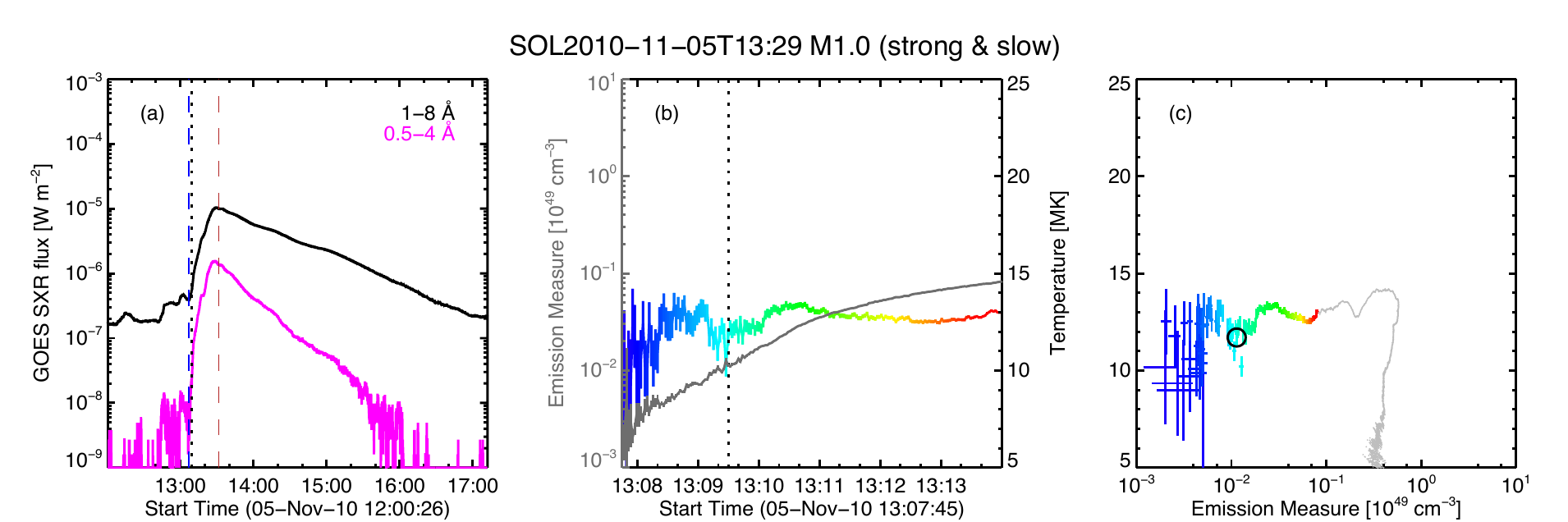}
\includegraphics[width=0.9\textwidth]{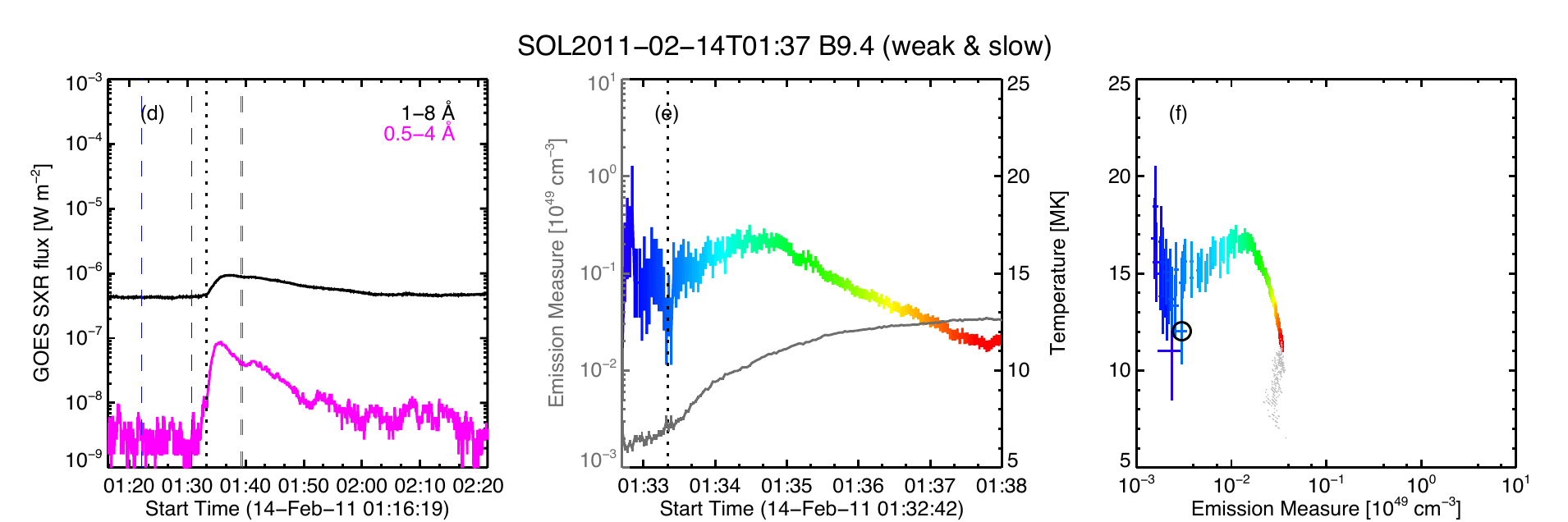}
\includegraphics[width=0.9\textwidth]{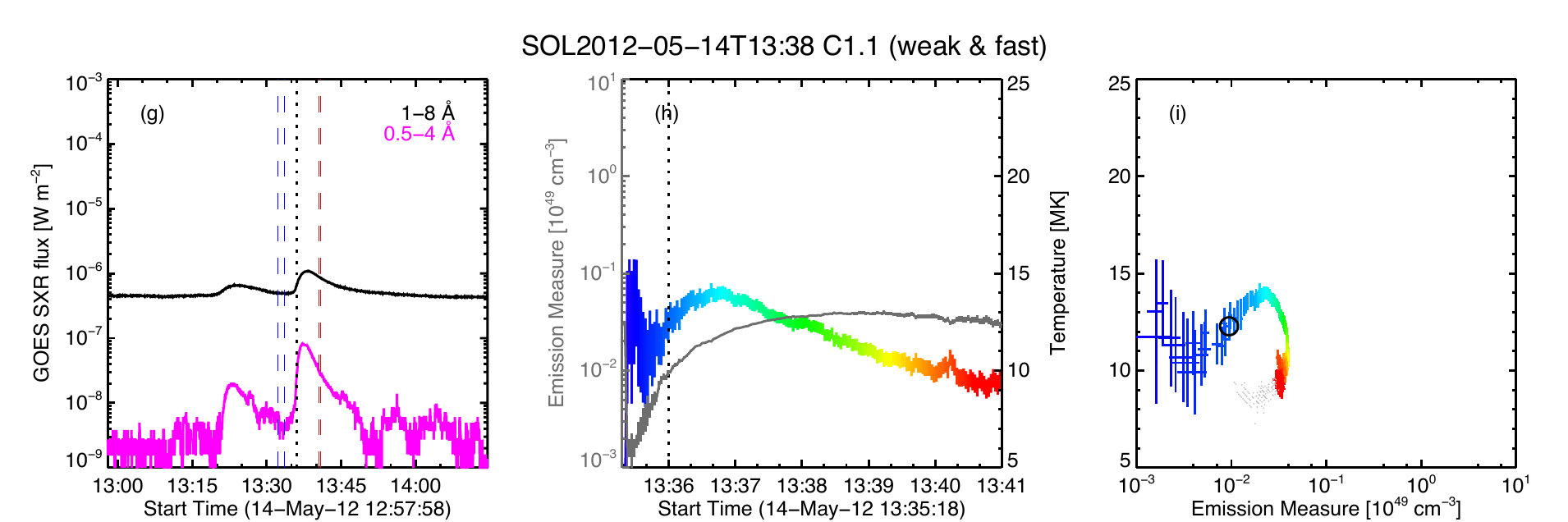}
\includegraphics[width=0.9\textwidth]{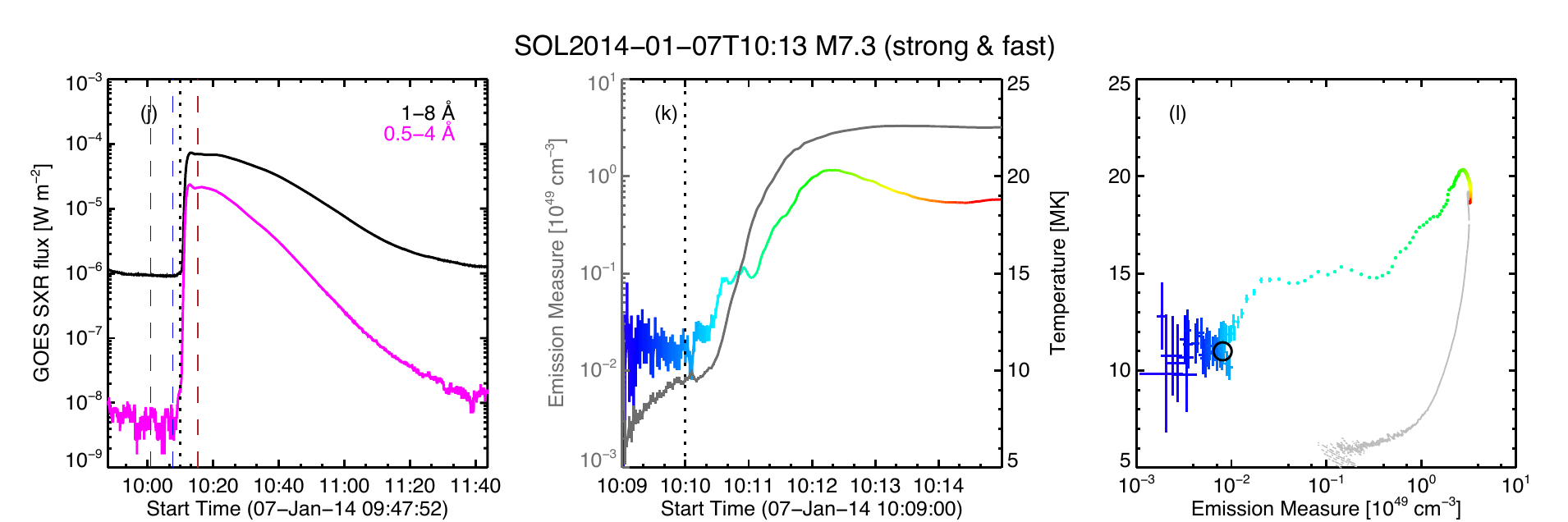}
\caption{GOES isothermal fit parameters for the four representative events. 
Left-hand column: GOES SXR flux data. The dashed lines show the time intervals for background (blue, see Table~\ref{tab:four}) and noise (red) estimation. The dotted vertical line shows the start time of the RHESSI HXR ($>$12~keV) emission; see Table~\ref{tab:temps}.
Center column: timeseries of the temperature (colour-coded) and the emission measure (grey). 
Right-hand column: correlation between temperature and emission measure. 
The hot onset times show up at the very beginning, at lowest emission measure and with temperatures in the 10-15~MK range. Note that the temperature uncertainties are overestimates, as described in the text. The black circles show the parameter state at the approximate time of the dotted lines in the left-hand column plots; 
The colour coding matches that of the timeseries plot.}
\label{fig:four_ts}
\end{figure*}

Figure~\ref{fig:four_ts} shows the GOES flux, temperature and emission measure results for four events (detailed in Table~\ref{tab:four}), chosen arbitrarily to represent fast, slow, strong, and weak flares respectively, crudely bracketing the parameter space of rise time and flare energy.
``Fast/slow'' refers to the event rise time and ``strong/weak'' refers to the GOES 1-8 \AA\ peak flux values. These typical events are from the 2011-2014 time frame and do not represent different flare classes as such, 
since the parameters generally have broad, continuous distributions
\citep[e.g.][]{1993ApJ...412..401L}. 

\begin{table*}
        \centering
        \caption{Four representative events}  
        \bigskip
        \label{tab:four}
        \begin{tabular}{llll} 
                \hline

                Flare (IAU) & GOES Background & GOES Onset   & Description  \\
                            & interval (UT)        & interval (UT)        &  \\

                 \hline
 \\

SOL2010-11-05T13:30 (M1.0) & 13:06:19 - 13:07:30 & 13:08:00 - 13:09:30 & strong, slow\\
SOL2011-02-14T01:37 (B9.4) & 01:22:09 - 01:30:47 & 01:32:00 - 01:33:20 & weak, slow \\
SOL2012-05-14T13:38 (C1.1) & 13:32:16 - 13:33:38 & 13:35:25 - 13:36:00 & weak, fast \\
SOL2014-01-07T10:13 (M7.3) & 10:00:57 - 10:07:36 & 10:09:05 - 10:10:00 & strong, fast\\

\hline
        \end{tabular}
\end{table*}

The hot onset sources appear substantially before the beginning of the impulsive phase (indicated by the dotted vertical lines in the left column panels in Fig.~\ref{fig:four_ts}), as determined by RHESSI hard X-rays above 25~keV where possible, and above 12~keV where not. 
The correlation between temperature and emission measure (right column in Figure~\ref{fig:four_ts}) shows a roughly clockwise circulation during the main phase of the flare,
ending with the cooling of the coronal loops.
The hot onset emission precedes these features, appearing at the lowest emission measure but an elevated temperature.
The cooling phase passes through the onset temperature range smoothly, establishing that the 10-15~MK level is not an artifact.
The data points are colour-coded and mapped to the temperature curves in the middle column of Figure~\ref{fig:four_ts} to indicate the time-evolution of this correlation. These panels indicate that the hot onset (with temperatures around $10\sim 15$ MK) is associated with a low amount of plasma, with emission measure values below $10^{47}$cm$^{-3}$. 
In the flare sample discussed here, the ``fast strong'' event SOL2014-01-07 (bottom row of Fig.~\ref{fig:four_ts}), for example, has a hot onset detectable more than a minute prior to the detectable HXR emission.
The GOES isothermal onset temperatures, \textit{i.e.} the first observable measurements lie well above the low-temperature range of the these data \citep[e.g.][]{1997ApJ...479L.149S,2005SoPh..227..231W}.
So far as the data permit us to tell, the first detected emission at these hot onset times already has a measurable temperature significantly above any observational limit.

\subsection{Uncertainties on GOES temperature measurements}

\begin{figure}
\centering
\includegraphics[width=0.45\textwidth]{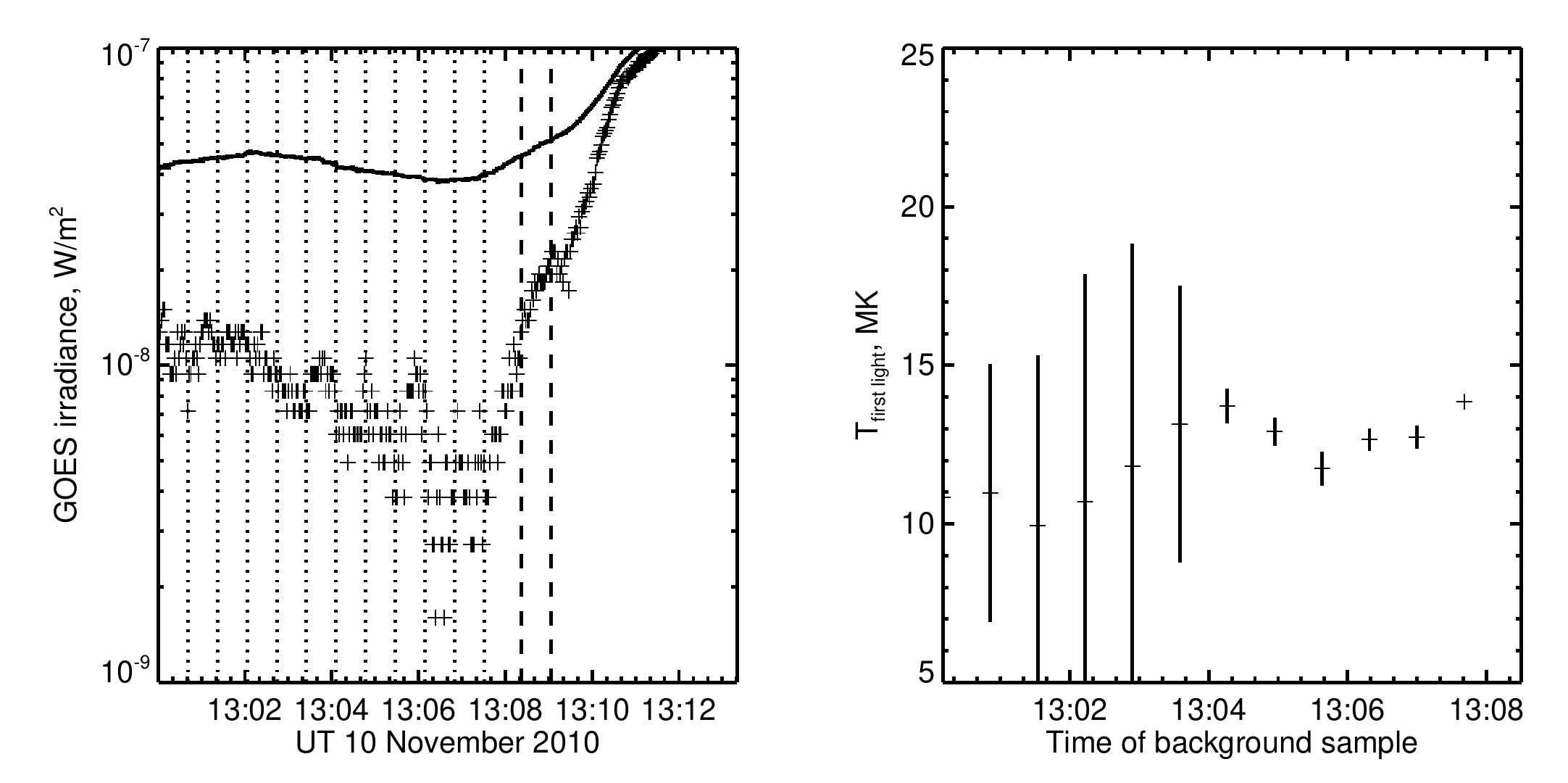}
\caption{illustration of the sensitivity of the hot onset temperature, at a fixed time interval, to different choices of background interval in the 
analysis of the GOES data for SOL2010-11-05.
Left, the two GOES time series (1-8 \AA\ divided by 10) with dotted lines showing different choices of background interval. 
Note the dominance of digitization error at low flux levels in the 0.5-4~\AA\ channel; right, the inferred temperature for the interval between the two dashed lines, plotted against the time at which each 20-s background interval was taken.}
\label{fig:background_sensitivity}
\end{figure}

The error bars in Figure~\ref{fig:four_ts} reflect both random errors, as estimated from the scatter of data at an intermediate flux level, and the digital uncertainty resulting from undersampling the true background noise, as discussed in \cite{2015SoPh..tmp...50S}.
The digital step size varies from epoch to epoch, since the different GOES satellites have different properties,
For GOES-15 in the 0.5-4~\AA\ channel it was $6 \times 10^{-10}$~W/m$^2$ equivalent, at the time of SOL2014-01-07 (peak 0.5-4~\AA\ flux $2.35 \times 10^{-5}$~W/m$^2$).
This is a crucial matter at flare onset under low-background conditions, because it often dominates the pre-flare fluctuation in the more important (in the sense of greater variance) 0.5-4~\AA\ band.
Figure~\ref{fig:background_sensitivity} illustrates the noise properties for one of our illustrative events.
During quiet conditions the 0.5-4~\AA\ background levels reflect intrinsic background may show only thermal noise and/or unwanted radiation effects rather than any solar source.
At these times the digital step size may exceed the intrinsic detector noise fluctuation, and this can add variance in a manner difficult to characterize in a transient.
Accordingly we have adopted the minimum digital increment as a noise term, directly added to the signal fluctuation determined via the RMS fluctuation at higher signal levels determined during a chosen interval at higher rates.
Here we take 10 data points one minute after the peak of the 1-8~\AA\ channel, as indicated by the red dashed vertical lines in the left-column panels of Figure~\ref{fig:four_ts}. 
We estimate the intrinsic detector noise fluctuation as the normalised standard deviation of the residuals of fitting this 10-datapoint interval with a 4th-order polynomial function, to which the digital noise is added linearly.
These noise estimates are then added to and subtracted from the background-subtracted GOES flux (on both channels) and fed into GOES\_TEM.pro to obtain the uncertainties for the temperature and emission measure values.

Figure~\ref{fig:background_sensitivity} shows the sensitivity of the inferred hot onset temperature determination to the choice of background time interval, for the case of SOL2010-11-05. Here the error bars just reflect the standard deviation of the 20-point sample at a fixed time interval, as shown by the dashed lines. The large error bars here show the observed standard deviations; the earliest background samples result in some non-physical flux ratios, for which the SolarSoft GOES\_TEM.pro algorithm returns a default 4~MK, and are thus data artifacts. Generally the closer in time to the measurement, the more trustworthy the background information should become.

\subsection{Time histories}

The two parameters derived from the isothermal GOES fits, temperature and emission measure, allow us to trace out an event's time evolution in a correlation plot of the temperature and emission measure parameters, a technique often used to illustrate the different flare phases, as shown in the right panels of Figure~\ref{fig:four_ts}.
The hot onset interval clearly stands out in each case, with little variation of temperature as the emission measure slowly grows.
For SOL2014-01-07 (M7.8) the emission measure of the hot onset source does not exceed 1\% of its ultimate maximum.
As expected, the start of the impulsive phase corresponds to a large increase in emission measure simultaneously with the presence of higher temperatures, in the pattern expected from the Neupert effect \citep{1967ApJ...149L..79N}. See for example Fig.~7 of \cite{2009A&A...494.1127R}.

The right-column panels in Figure~\ref{fig:four_ts} show only the beginning of each event with a colour-coded time for clarity.
We also note that the completion of the trajectory in grey shows that the cooling branch of the diagram has no peculiarities in the temperature range of the hot onset. The late phase, densely sampled because of its relatively slow evolution, shows the pattern expected of ``overdense'' flare loops cooling as result of the cessation of major energy release \citep{1991A&A...241..197S,1995ApJ...439.1034C,2009A&A...494.1127R}.

\subsection{EUV/UV Images}\label{sec:images}

\begin{figure*}
\centering
\includegraphics[width=0.9\textwidth]{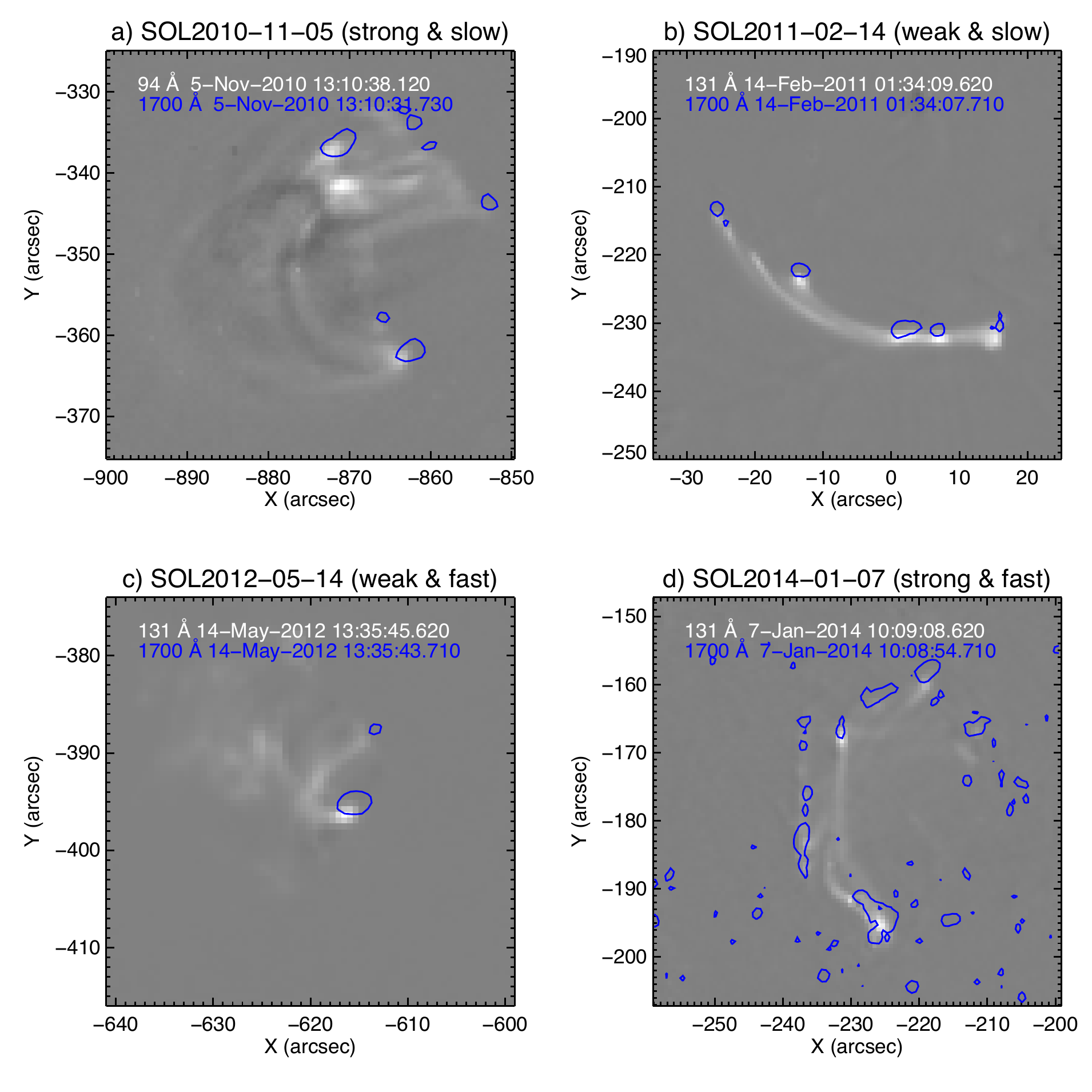}
\caption{AIA difference images for a hot channel (94 or 131~\AA), overlaid with 1700~\AA\ contours at 20\% of the image maximum taken from a running-difference image. (a) SOL2010-11-05, base image at 13:04:02 UT. (b) SOL2011-02-14, base image at 01:28:09 UT. (c) SOL2012-04-14, base image at 13:33:09 UT. (d) SOL2014-01-07, base image at 10:06:20 UT.}
\label{fig:four}
\end{figure*}

The GOES timeseries (Figure~\ref{fig:four_ts}) identify the time range of the hot onset interval for each of the four representative flares, and we identify these spatially via the corresponding AIA 94 or 131~\AA\ images in Figure~\ref{fig:four}. In each panel we show a difference image between the onset time and the background time and then overlay with contours from the closest AIA 1700~\AA\ images to the onset time, highlighting the chromospheric flare emission \citep{2019ApJ...870..114S}. The close association of the bright and localised sources from a high temperature plasma (94 or 131~\AA) and chromospheric footpoints (1700~\AA\ contours) is strong evidence for the presence of plasma of temperatures around 10~MK at the flare footpoints even at these early stages of flares. As expected, the images of the two ``fast'' events (upper left, lower right) show that they are smaller physically. The AIA images were processed using the standard software \citep{2012SoPh..275...41B}.

These images confirm that the onset source regions include footpoints, identifiable with subsequent flare footpoint sources. 
Some loop emission also appears.
One of the three cases shows a simple double footpoint pattern, but the others have multiple regions appearing simultaneously in these single snapshot images at exposure times of 2.9~s.

\begin{figure*}
\centering
\includegraphics[width=0.9\textwidth]{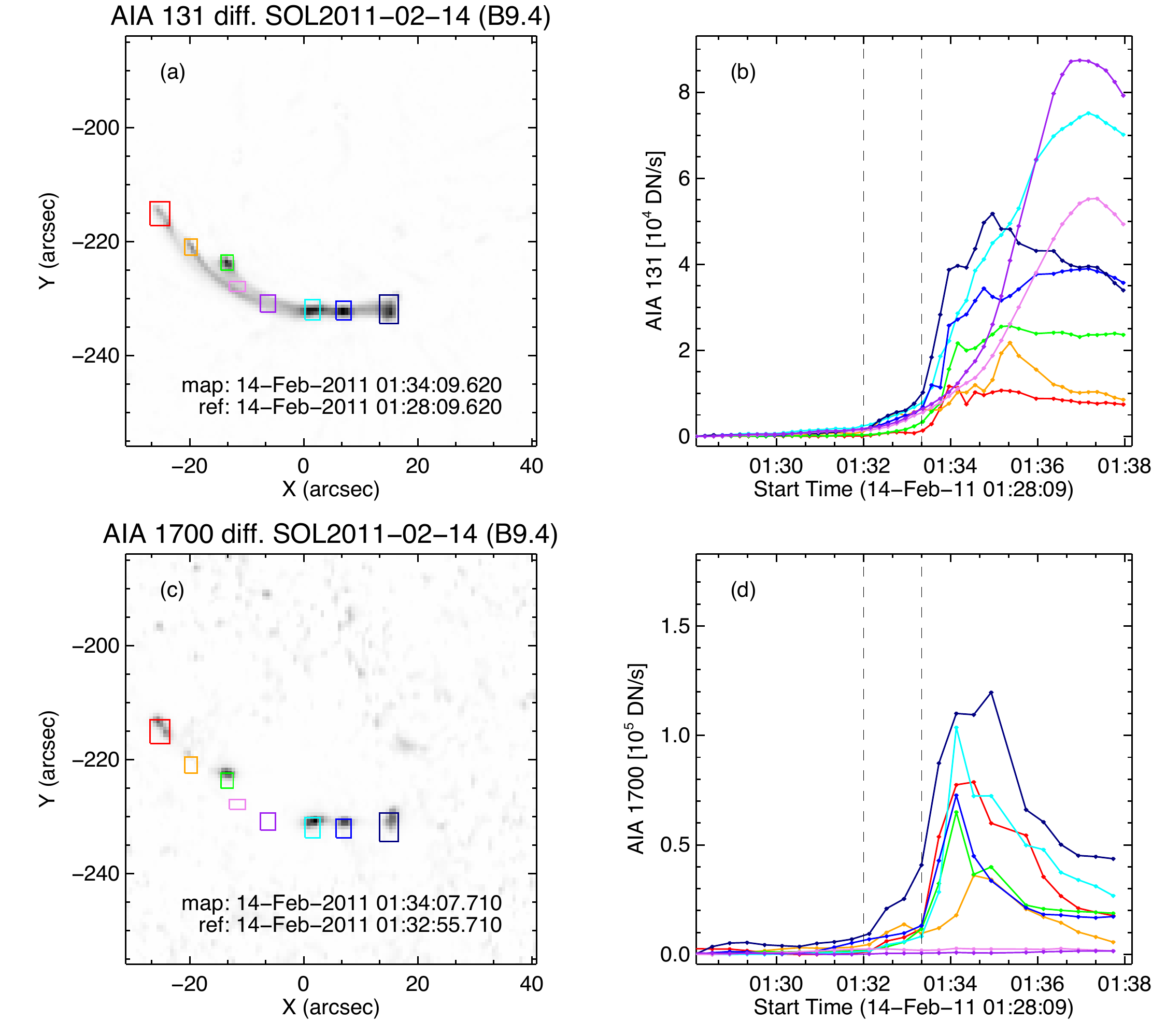}
\caption{Identification of footpoints from AIA 131~\AA~(top) and 1700~\AA~images (bottom), with images on the left and their relative flux contributions (right). 
The dashed vertical lines show the onset interval, identified from GOES SXR data (see Table~\ref{tab:four}).}
\label{fig:fpimage}
\end{figure*}

\begin{figure}
\centering
\includegraphics[width=0.5\textwidth]{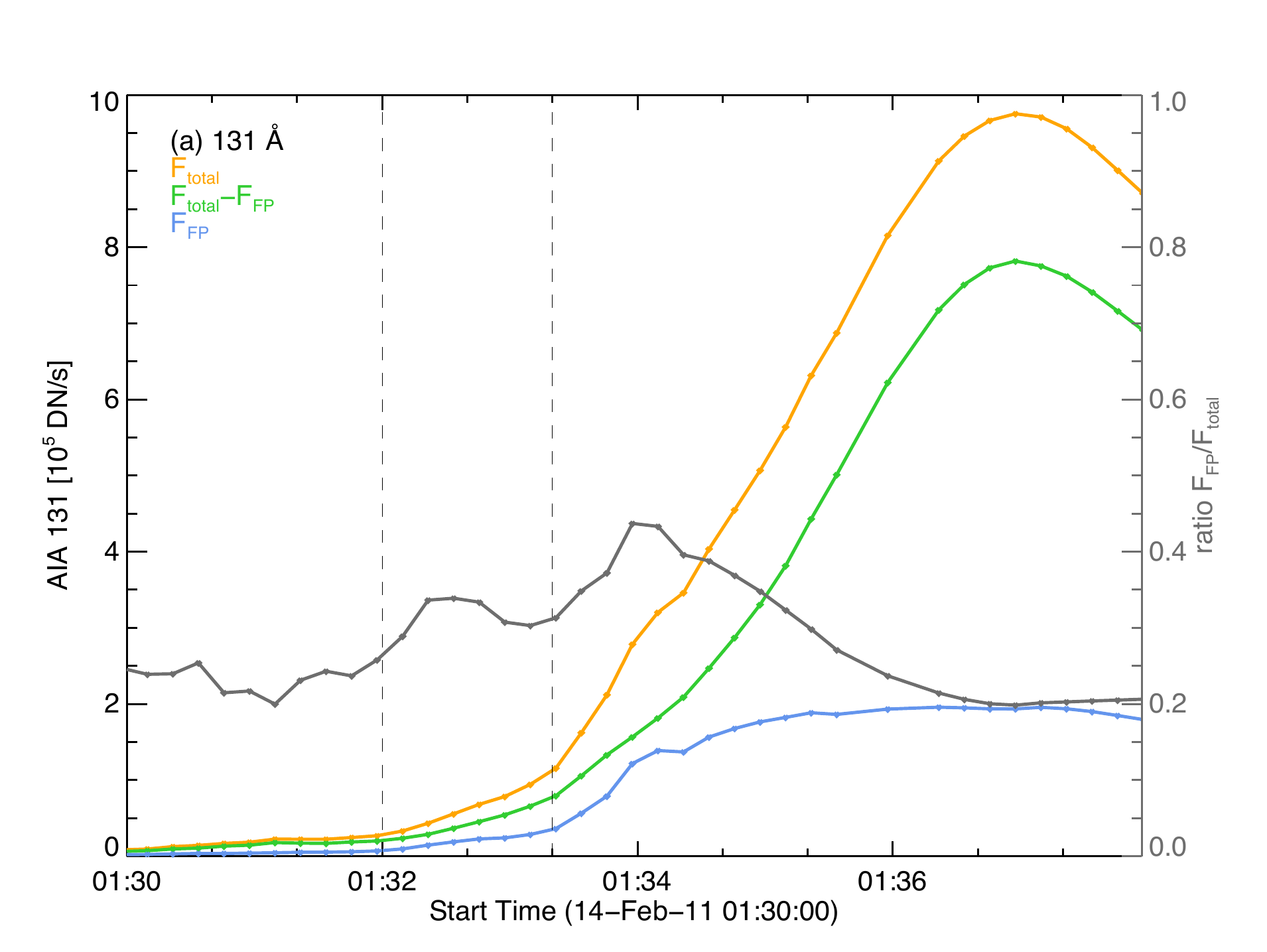}
\includegraphics[width=0.5\textwidth]{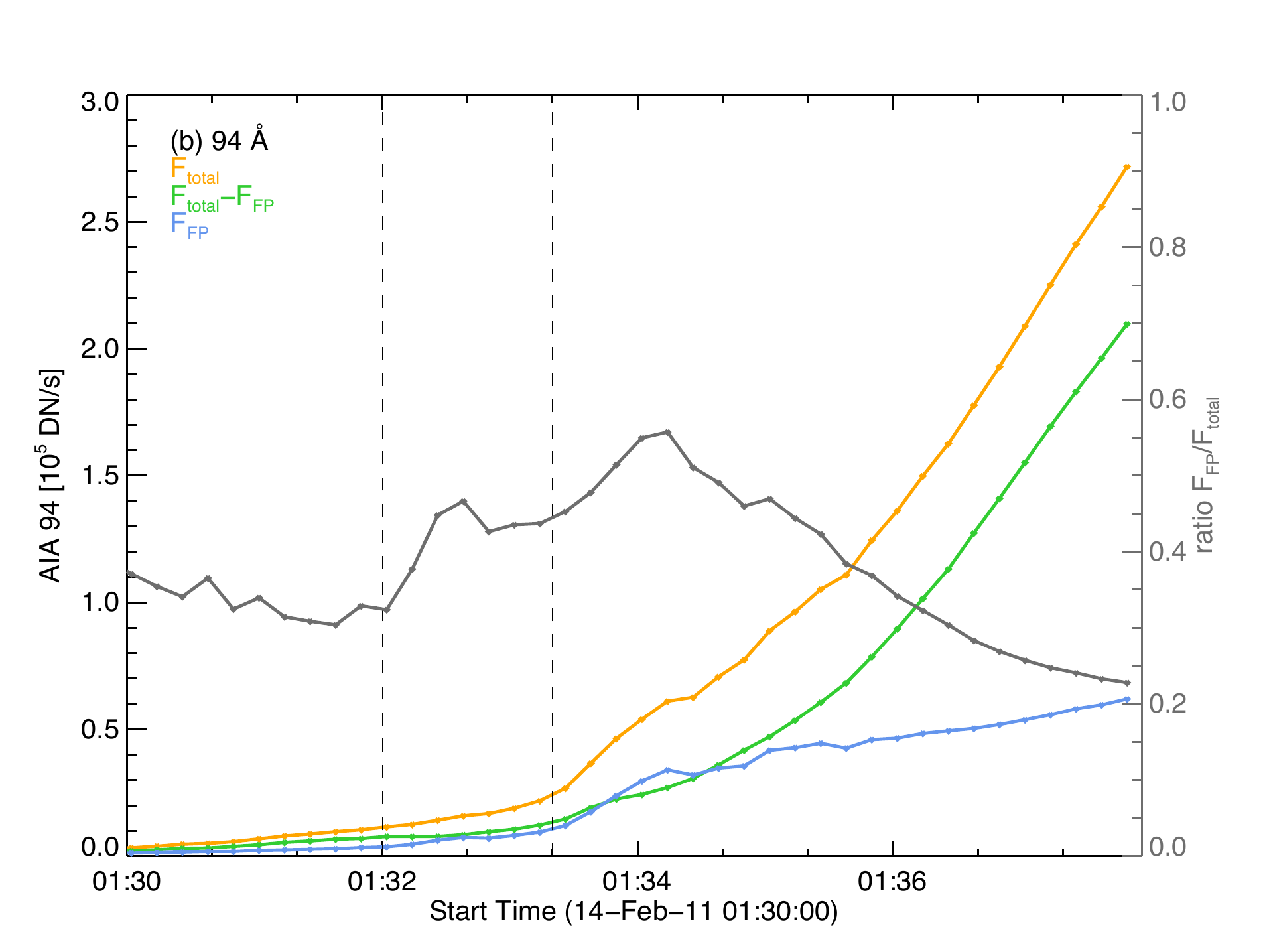}
\caption{AIA 131~\AA\ and 94~\AA\ total flux (yellow), flux in the boxed footpoint sources regions defined in Figure~\ref{fig:fpimage} (blue), and the difference between them as an estimate of the emission from the coronal loops (green). 
The ratio of footpoint flux to total flux is shown in black. The dashed vertical lines show the onset interval, identified from GOES SXR data (see Table~\ref{tab:four}).} 
\label{fig:fp}
\end{figure}

Figures~\ref{fig:fpimage} and \ref{fig:fp} examine the time variations of image features for SOL2011-02-14, in order to quantify the EUV fluxes in the footpoints relative to the main body of the flare. Note the presence of multiple compact sources in this event, in the hot onset interval. These identify well with the footpoints of faint loop features, but do not show an orderly ribbon organization. 
The clear identification of the footpoints in both spectral bands suggests broad contributions from multithermal plasmas \citep[e.g. ][]{2015SoPh..290.3573S}. 
The EUV 131~\AA\ and UV 1700~\AA\ intensity of each identified footpoint, indicated by the coloured boxes, is summed to construct the flux timeseries. In Figure~\ref{fig:fpimage}, the pink and purple boxes mark regions of the coronal loops connecting the footpoints; their 131~\AA\ fluxes show the typical gradual rise originated from the hot plasma filling these loops, which display no emission at UV 1700\AA. 
On the other hand, the footpoints (marked by the rest of the coloured boxes) do show the emission rising in both 131~\AA\ and 1700~\AA\ as early as 01:32:00~UT, with both having a more impulsive development, reaching their maximum values after 01:34~UT, very close to the maximum of the impulsive phase of this event as indicated by the peak of the HXR emission. 

These comparisons confirm, in a straightforward manner, that 10-15~MK temperatures appear in the footpoint sources simultaneously with the `typical' chromospheric temperatures.

Quantitative analysis of the source fluxes in terms of their physical parameters lies outside the scope of this work, but we can readily compute the ratios of emission from footpoint and loop regions in the two bands, as shown in Figure~\ref{fig:fp}. 
In this figure, $F_\mathrm{total}$ represents the flux of the entire region's field-of-view (Figure~\ref{fig:fpimage}a), $F_\mathrm{FP}$ is the total flux from all the footpoint boxes in Figure~\ref{fig:fpimage}a, and $F_\mathrm{total}-F_\mathrm{FP}$ gives an estimate of the emission originating only from the loops. 
The ratio $F_\mathrm{FP}/F_\mathrm{total}$ (in grey) confirms the relatively strong contributions (at the 20\% - 50\% level) from the footpoint regions in the hot onset interval.

\section{Confirmation of the hot onset property}

\subsection{Occulted flares}

\begin{figure}
\centering
\includegraphics[width=0.45\textwidth]{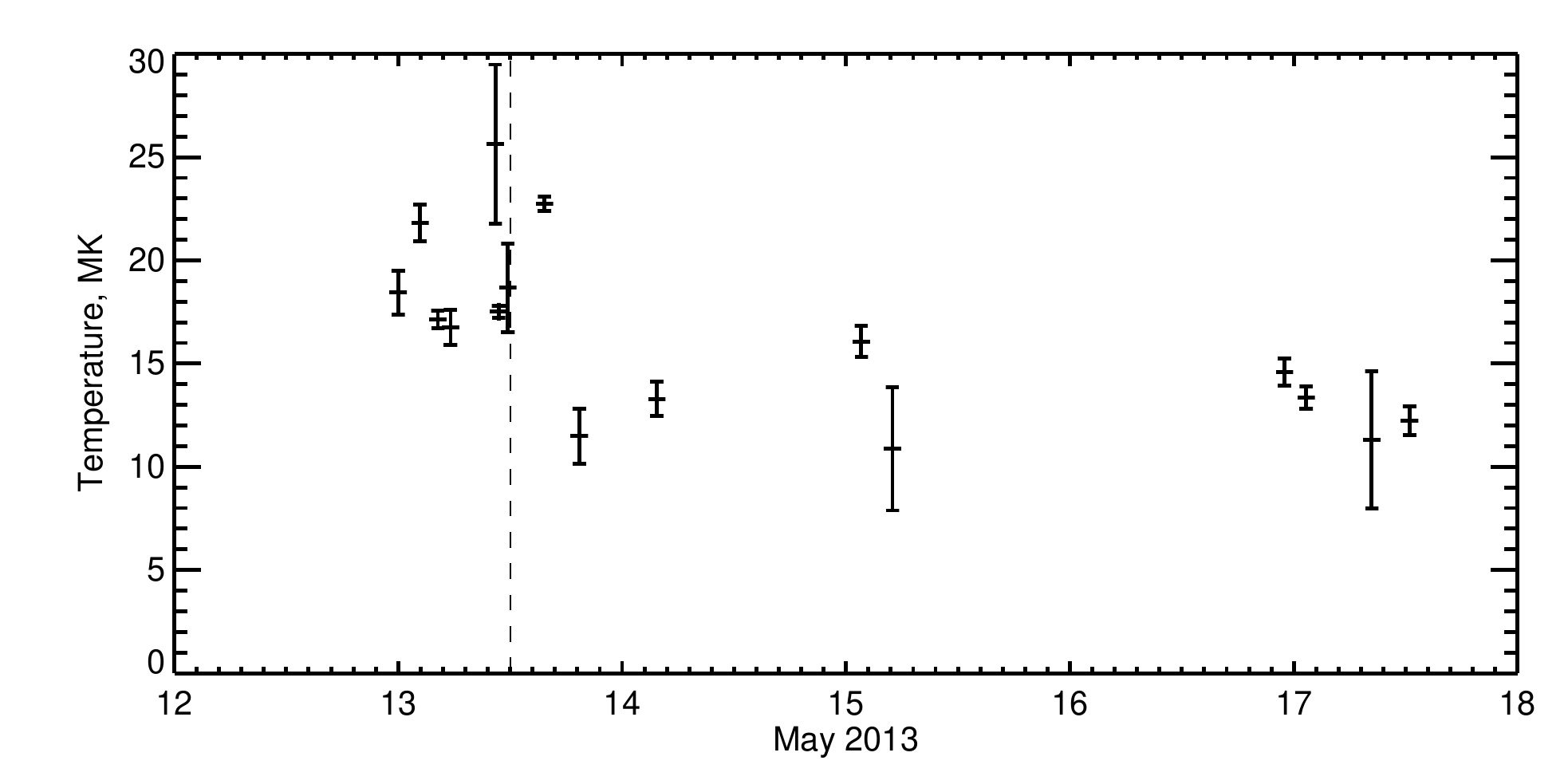}
\caption{Earliest GOES temperatures of flares measured during the east limb transit of NOAA AR 11748 in 2013, with the approximate time (mid-day May 13) of the transit indicated by a vertical dashed line.
The higher temperatures for the occulted flares are consistent with the higher temperatures of flare arcades on larger spatial scales, whereas the 10-15~MK onset
sources have been occulted.
Note that this plot omits SOL2013-05-14 (X3.2) because of background confusion.}
\label{fig:occ_plot}
\end{figure}

The image analysis in Section~\ref{sec:images} shows that we can identify the hot onset X-ray sources with corresponding AIA 1700~\AA\ features, consistent with chromospheric heights rather than a coronal origin.
For a flare with footpoints occulted by the limb, just prior to or just after a limb passage, we would expect different properties.
The footpoint regions in such occulted flares cannot be seen and the SXR emission must therefore come from large-scale loops (the flare arcade).
In major events these tend to show temperatures higher than the 10-15~MK hot onset range \citep{1994SoPh..154..275G}.
We confirm this in Figure~\ref{fig:occ_plot}, which shows how the onset temperature varies as NOAA active region 11748 transited the east limb in May 2013. The coronal sources seen before limb transit (approximately mid-day May 2013) have temperatures in the 15-25~MK range, depending on flare magnitude, while the on-disk sources seen after the transit tend to have temperatures in the 10-15~MK range. This sample also helps to confirm the universality of the hot onset phenomenon since it consists of additional events selected without bias.

\subsection{RHESSI observations}\label{sec:rhessi}

\begin{figure*}
\centering
\includegraphics[width=0.49\textwidth]{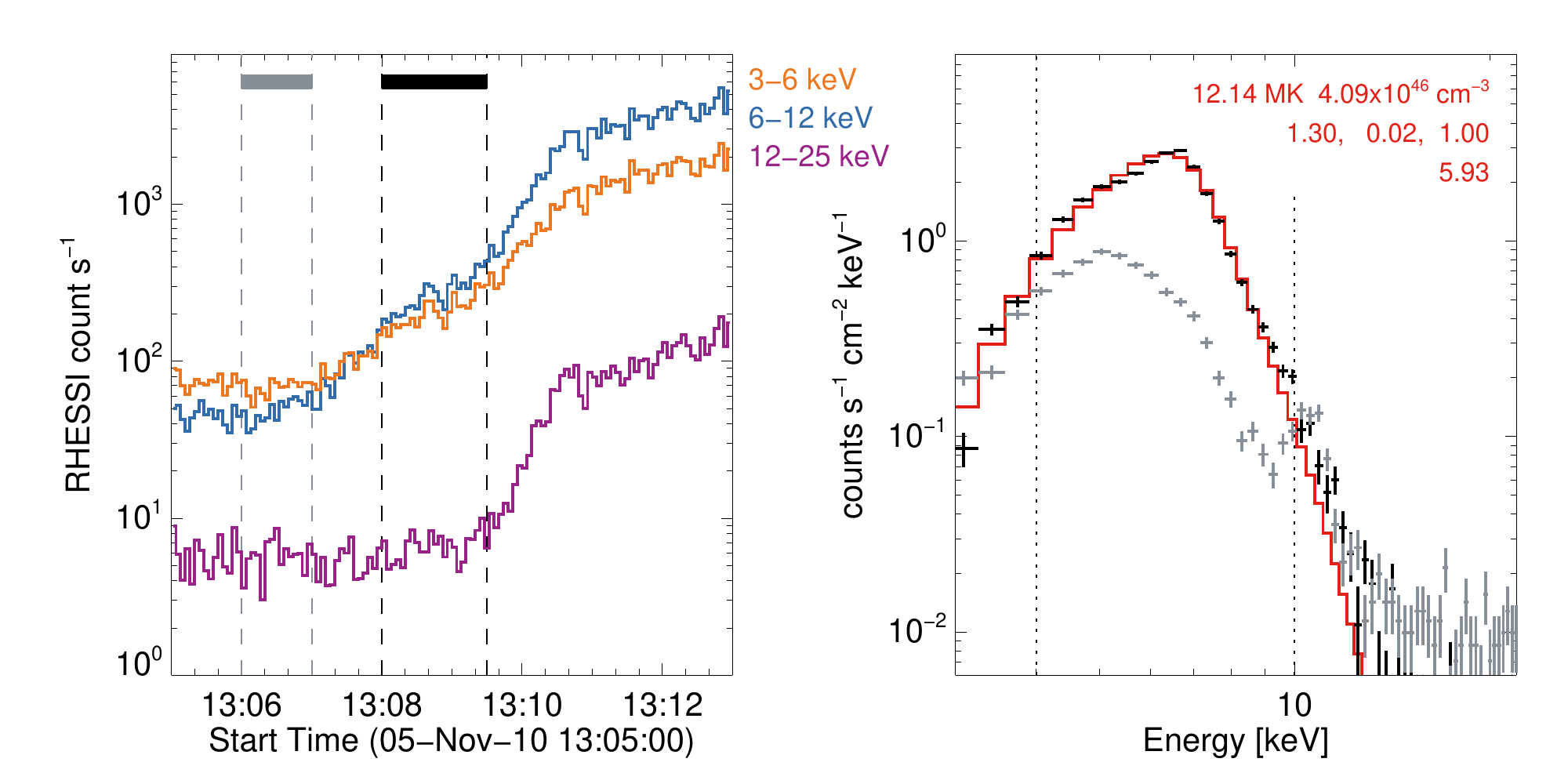}
\includegraphics[width=0.49\textwidth]{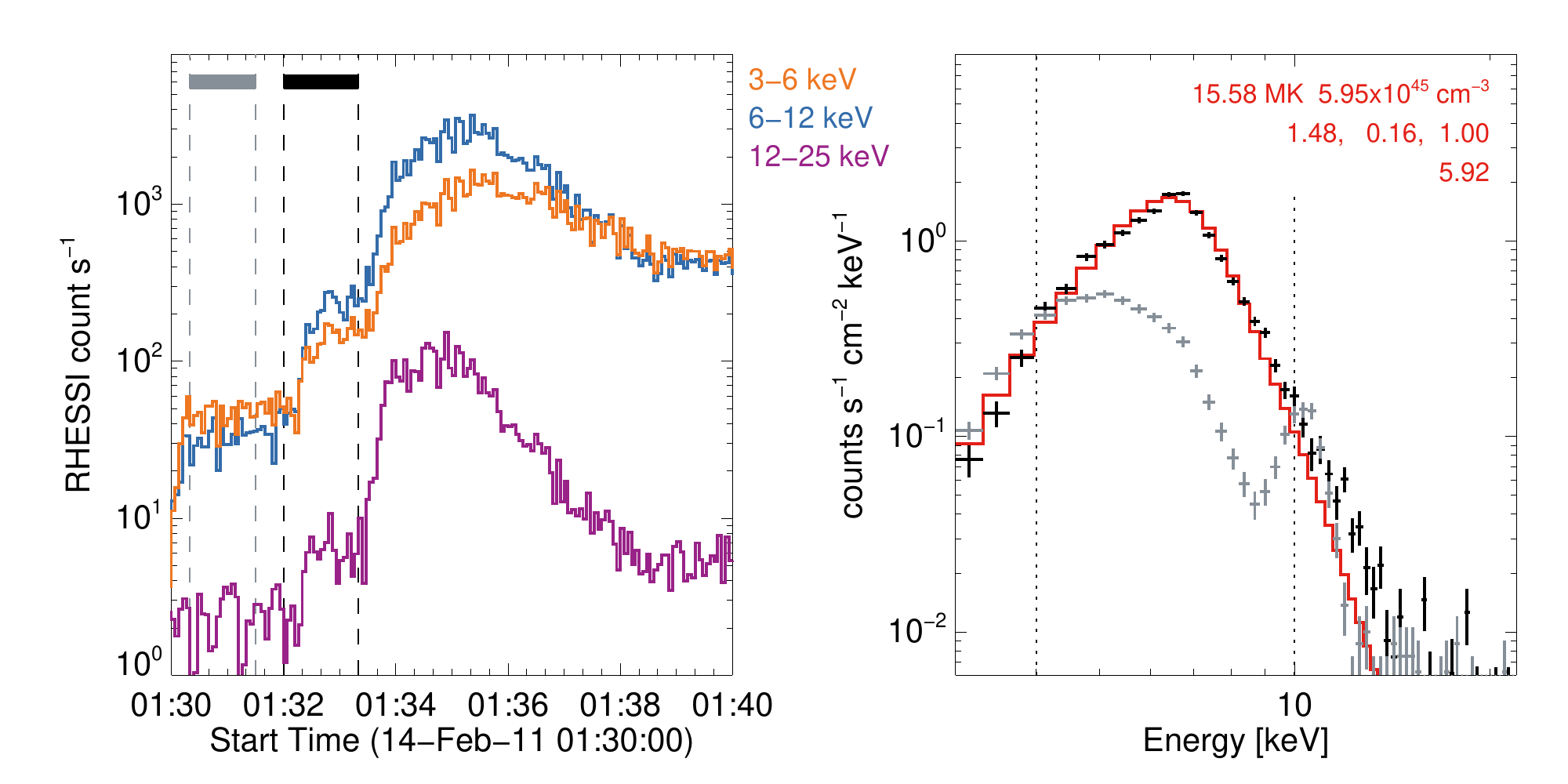}
\includegraphics[width=0.49\textwidth]{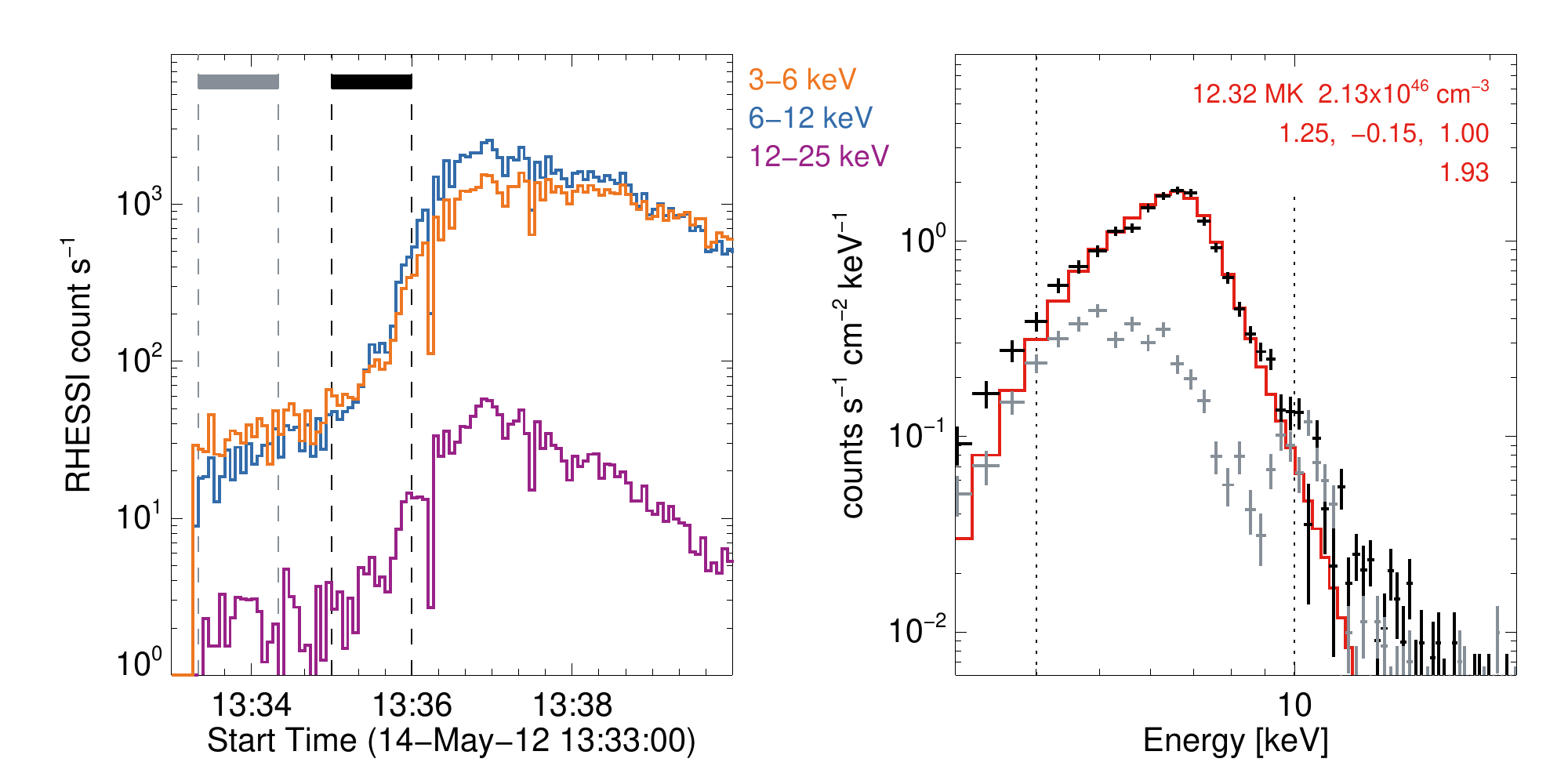}
\includegraphics[width=0.49\textwidth]{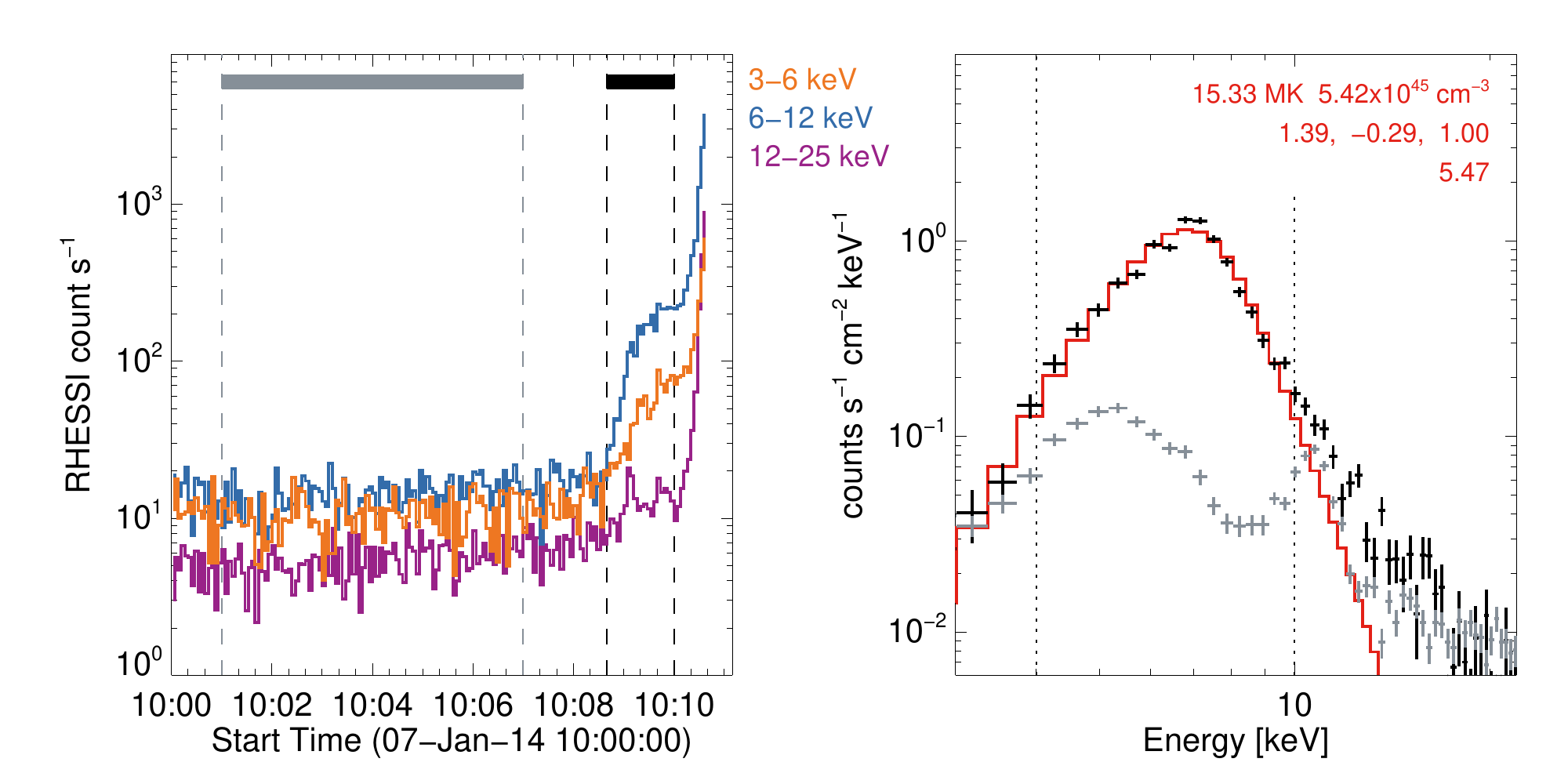}
\caption{RHESSI lightcurves and fitted spectra for each of the four flares. The lightcurves show the RHESSI emission in 3 different energy bands, and indicate the time range used for the background (grey dashed vertical lines and bar) and onset (black dashed vertical lines and bar) spectra. The spectra for the background time (grey data points) and background subtracted data (black points) are shown and fitted with an isothermal model (red, parameters given) over 4-10 keV (blacked dotted vertical line).}
\label{fig:rhessi}
\end{figure*}

The RHESSI X-ray data provide a check on the temperatures obtained from GOES. 
We have analyzed these observations for the four sample events, as shown in Figure~\ref{fig:rhessi}, where we show the lightcurves in 3-6, 6-12 and 12-25 keV for each event along with a fitted spectrum. 
The lightcurve for the highest energy band (12-25~keV) typically has a non-thermal interpretation, with the lower energy bands reflecting predominantly thermal emission. 
The lightcurves can be shown in full for the SOL2011-02-14 and SOL2-12-05-14 flares, but for the larger flares (SOL2010-11-05 and SOL2014-01-07) we only show the initial period before RHESSI's attenuating shutter comes in (after the start of HXR emission in both cases). 
In all events there is a steady increase in the thermal lightcurves during the GOES hot onset time, and it is over this time that we fit a RHESSI X-ray spectrum. 
In Figure~\ref{fig:rhessi} we show the background spectrum from a pre-onset time, and the onset spectrum with this background subtracted.
The spectral fit uses the {\textit {f\_vth.pro}} thermal model in the Object Spectral Executive \citep[OSPEX software, ][]{2002SoPh..210..165S}. 
The resulting fit parameters and uncertainties are given in Figure~\ref{fig:rhessi} and Table~\ref{tab:temps}. 
In each case the RHESSI spectrum shows a temperature between 12 - 15~MK, consistent with the GOES temperatures derived for these flares. 
Added to this thermal model is the {\textit {drm\_mod.pro}} pseudo function which takes care of the degraded detector performance (increased noise and poorer energy resolution due to radiation damage) which is present during these flares. 
This also requires the analysis to be performed for single RHESSI detectors; the data shown in Figure~\ref{fig:rhessi} is for one of the better-performing detector during each flare, which is detector~9 for the first three events, but detector~1 during the last event. 
Several of the other detectors produce similar, if noisier results. 
In some of the flare spectra shown in Figure~\ref{fig:rhessi} there is a slight excess in counts over the model $>10$~keV, which might be indicative of very weak HXR emission, but in each case this signal is less than $2\sigma$ above the background. 
The strongest case of this is for SOL2014-01-07, but RHESSI degradation was relatively severe at the time of this event. 
This degradation is also the likely reason for the poorer consistency between GOES and RHESSI temperatures in this event compared to the other flares.

\begin{table*}
    \centering
    \caption{GOES/RHESSI temperature comparisons}
    \bigskip
    \label{tab:temps}
        \begin{tabular}{llllllll} 
                \hline
                Flare (IAU) & RHESSI Background & RHESSI Onset &  \multicolumn{2}{c}{Temperature [MK]}   \\
                             & interval (UT)  & interval (UT)  & GOES & RHESSI   \\
                 \hline 
SOL2010-11-05T13:30 & 13:06:00 - 13:07:00 & 13:08:00 - 13:09:30 & 12.2 $\pm$ 1.1  & 12.14 $\pm$  0.52  &&  \\
SOL2011-02-14T01:37 & 01:30:20 - 01:31:30 & 01:32:00 - 01:33:20 & 14.5 $\pm$ 1.7  & 15.58 $\pm$  1.15  && \\
SOL2012-05-14T13:38 & 13:33:20 - 13:34:20 & 13:35:00 - 13:36:00 & 11.6 $\pm$ 0.9  & 12.32 $\pm$  0.49  && \\
SOL2014-01-07T10:13 & 10:01:00 - 10:07:00 & 10:08:40 - 10:10:00 & 11.1 $\pm$ 0.6  & 15.33 $\pm$  1.85  && \\
        \hline
        \end{tabular}
\end{table*}

\subsection{Summary}

The observations described above can be summarized briefly as follows:
\begin{enumerate}
    \item The soft X-ray onsets of solar flares detected by GOES/XRS tend to have elevated temperatures already at the time of initial detectability.
    \item The temperatures of these sources range from 10-15~MK and do not depend strongly on flare magnitude or configuration.
    \item These ``hot onset'' sources lie in the lower atmosphere, including multiple simultaneous footpoint brightenings that appear prior to the impulsive phase and make a substantial contribution to the overall high-temperature signature.
    \item The onset high temperatures appear within 10~s of the flare's detectable onset in the GOES 0.5-4~\AA\ channel, and vary only slowly over the hot onset interval, for a duration of 10s of seconds to minutes.
    \item We have confirmed the GOES results with RHESSI data.
\end{enumerate}

\section{Discussion}

What physical conditions prevail in the hot onset sources?
A systematic study is beyond the scope here, but we can comment on the densities obtained by comparing the emission measures (Figure~\ref{fig:four_ts}) with the image scales seen in Figure~\ref{fig:four}.
A typical emission measure for a hot onset footpoint source is about 20\% to 40\% (Figure~\ref{fig:fp})  of $n_e n_i V = 10^{47}$~cm$^{-3}$ (Fig.~\ref{fig:four_ts}), where $n_e$ and $n_i$ are the electron and ion densities, respectively.
Assuming $n_e = n_i$ and volume $V = 10^{8} \times 10^8 \times 10^7$~cm$^{3}$ (based on typical sizes of the hot footpoints), one finds $(3.0 < n < 4.7) \times 10^{11}$~cm$^{3}$, consistent with a chromospheric region. 
The SolarSoft temperature estimates, both for GOES and RHESSI data, assume collisional ionization equilibrium \citep[e.g.][]{2009A&A...502..409B}. 
This may be a reasonable assumption for these hot, dense chromospheric sources, in the absence of evidence for a significant role for collisional excitation by non-thermal electrons. 
However, a full exploration with a dynamic model capable of tracking the ionization states in detail should follow; in particular the line-to-continuum ratio may differ considerably in regions of partial ionization.

\section{Conclusions}

Our work with the GOES soft X-ray observations in the earliest detectable stages of solar flares has revealed an unexpectedly common temperature behavior, with flares of all magnitudes starting at temperatures of 10-15~MK with no sign of gradual temperature increase.
We have identified these earliest source regions mainly as footpoints and low-lying loops that become part of the structure of the flare.
The behavior of these sources may explain some of the features of the frequently-reported flare precursors as seen in other ways.
They offer a substantial challenge (and an opportunity) for standard modeling techniques.
The recognition of the hot onset phenomenon immediately challenges the modeling as carried out with 1D radiation hydrodynamics in the standard framework \citep[][\textit{et seq.}]{1980SoPh...68..351N}. 
So far as we are aware this modeling approach has not actually predicted the behavior that we observe.
We suggest that identifying the physical parameters responsible for the observed pattern should be a primary task for future work with these models,
including consideration of Alfv{\'e}nic energy transport \citep{2008ApJ...675.1645F}.
The hot onset sources have diagnostic importance for the models simply because they occur first, and avoid thus avoid confusion with the subsequent development of coronal flare loops.
We emphasize that the hot onset emission precedes the impulsive-phase ``hot footpoint'' phenomenon and differs from it \citep{1993ApJ...416L..91M,1994ApJ...422L..25H,2004A&A...415..377M,2013ApJ...771..104F,2013ApJ...767...83G,2015SoPh..290.3573S}.
We do not think that the ubiquitous nature of this phenomenon has been previously reported, but remark that the signatures certainly have appeared in individual evens reported earlier, even if not noted as having any generality, by \cite[e.g.][]{1985ApJ...298..887C} and \cite{2011ApJ...733...37F}.
The latter indeed drew the conclusion that a standard electron-beam model could fully explain all of the observed parameter evolution if the low-energy cutoff parameter is allowed to vary freely, but our findings disagree with that. Because of the absence of observable HXR emission, this phenomenon clearly represents a flare heating process that is physically different from that of the impulsive phase. 
We have confirmed that RHESSI observations also show these early high temperaures and look forward to a more detailed analysis of the hard X-ray properties, either with RHESSI or with more sensitive focusing optics, such as those pioneered by \textit{Focusing Optics X-ray Solar Imager} \citep[FOXSI, ][]{2009SPIE.7437E..05K} and NuSTAR \citep{2016ApJ...826...20G}, or especially with future missions such as \textit{Fundamentals of Impulsive Energy Release in the Corona Explorer} \citep[FIERCE, ][]{2019AGUFMSH33A..08S}.

\section{Data availability}

All data used in this paper reside in the public domain: for example for GOES, access via \url{https://www.ngdc.noaa.gov/stp/satellite/goes/index.html}; RHESSI via \url{https://hesperia.gsfc.nasa.gov/ssw/hessi/doc/guides/hessi_data_access.htm}, and SDO/AIA via \url{https://sdo.gsfc.nasa.gov/data/}.

\section{acknowledgments}

We thank NASA's Richard Schwartz and Kim Tolbert for maintaining the GOSS/XRS archive in such a scientifically useful format. PJAS acknowledges support from the Fundo de Pesquisa Mackenzie (MackPesquisa), CNPq (contract 307612/2019-8) and FAPESP (contract 2013/24155-3). IGH is supported by a Royal Society University Fellowship. LF acknowledges support from UK Research and Innovation's Science and Technology Facilities Council under grant award numbers ST/P000533/1 and ST/T000422/1. LAH is supported by an appointment to the NASA Postdoctoral Program at Goddard Space Flight Center, administered by USRA through a contract with NASA. HSH expresses thanks to the University of Glasgow for hospitality.

\bibliographystyle{mnras}
\bibliography{firstlight}

\begin{thebibliography}{}
\makeatletter
\relax
\def\mn@urlcharsother{\let\do\@makeother \do\$\do\&\do\#\do\^\do\_\do\%\do\~}
\def\mn@doi{\begingroup\mn@urlcharsother \@ifnextchar [ {\mn@doi@}
  {\mn@doi@[]}}
\def\mn@doi@[#1]#2{\def\@tempa{#1}\ifx\@tempa\@empty \href
  {http://dx.doi.org/#2} {doi:#2}\else \href {http://dx.doi.org/#2} {#1}\fi
  \endgroup}
\def\mn@eprint#1#2{\mn@eprint@#1:#2::\@nil}
\def\mn@eprint@arXiv#1{\href {http://arxiv.org/abs/#1} {{\tt arXiv:#1}}}
\def\mn@eprint@dblp#1{\href {http://dblp.uni-trier.de/rec/bibtex/#1.xml}
  {dblp:#1}}
\def\mn@eprint@#1:#2:#3:#4\@nil{\def\@tempa {#1}\def\@tempb {#2}\def\@tempc
  {#3}\ifx \@tempc \@empty \let \@tempc \@tempb \let \@tempb \@tempa \fi \ifx
  \@tempb \@empty \def\@tempb {arXiv}\fi \@ifundefined
  {mn@eprint@\@tempb}{\@tempb:\@tempc}{\expandafter \expandafter \csname
  mn@eprint@\@tempb\endcsname \expandafter{\@tempc}}}

\bibitem[\protect\citeauthoryear{{Awasthi} \& {Jain}}{{Awasthi} \&
  {Jain}}{2011}]{2011ASInC...2..297A}
{Awasthi} A.~K.,  {Jain} R.,  2011, in Astronomical Society of India Conference
  Series. pp 297--305

\bibitem[\protect\citeauthoryear{{Boerner} et~al.,}{{Boerner}
  et~al.}{2012}]{2012SoPh..275...41B}
{Boerner} P.,  et~al., 2012, \mn@doi [\solphys] {10.1007/s11207-011-9804-8},
  \href {https://ui.adsabs.harvard.edu/abs/2012SoPh..275...41B} {275, 41}

\bibitem[\protect\citeauthoryear{{Bradshaw}}{{Bradshaw}}{2009}]{2009A&A...502..409B}
{Bradshaw} S.~J.,  2009, \mn@doi [\aap] {10.1051/0004-6361/200810735}, \href
  {http://adsabs.harvard.edu/abs/2009A%26A...502..409B} {502, 409}

\bibitem[\protect\citeauthoryear{{Cargill}, {Mariska}  \&
  {Antiochos}}{{Cargill} et~al.}{1995}]{1995ApJ...439.1034C}
{Cargill} P.~J.,  {Mariska} J.~T.,   {Antiochos} S.~K.,  1995, \mn@doi [\apj]
  {10.1086/175240}, \href
  {https://ui.adsabs.harvard.edu/abs/1995ApJ...439.1034C} {439, 1034}

\bibitem[\protect\citeauthoryear{{Cheng}, {Pallavicini}, {Acton}  \&
  {Tandberg-Hanssen}}{{Cheng} et~al.}{1985}]{1985ApJ...298..887C}
{Cheng} C.~C.,  {Pallavicini} R.,  {Acton} L.~W.,   {Tandberg-Hanssen} E.,
  1985, \mn@doi [\apj] {10.1086/163672}, \href
  {https://ui.adsabs.harvard.edu/abs/1985ApJ...298..887C} {298, 887}

\bibitem[\protect\citeauthoryear{{Dere}, {Horan}  \& {Kreplin}}{{Dere}
  et~al.}{1974}]{1974JATP...36..989D}
{Dere} K.~P.,  {Horan} D.~M.,   {Kreplin} R.~W.,  1974, \mn@doi [Journal of
  Atmospheric and Terrestrial Physics] {10.1016/0021-9169(74)90008-7}, \href
  {https://ui.adsabs.harvard.edu/abs/1974JATP...36..989D} {36, 989}

\bibitem[\protect\citeauthoryear{{Dere}, {Landi}, {Mason}, {Monsignori Fossi}
  \& {Young}}{{Dere} et~al.}{1997}]{1997A&AS..125..149D}
{Dere} K.~P.,  {Landi} E.,  {Mason} H.~E.,  {Monsignori Fossi} B.~C.,   {Young}
  P.~R.,  1997, \mn@doi [\aaps] {10.1051/aas:1997368}, \href
  {http://adsabs.harvard.edu/abs/1997A%26AS..125..149D} {125, 149}

\bibitem[\protect\citeauthoryear{{Falewicz}, {Siarkowski}  \&
  {Rudawy}}{{Falewicz} et~al.}{2011}]{2011ApJ...733...37F}
{Falewicz} R.,  {Siarkowski} M.,   {Rudawy} P.,  2011, \mn@doi [\apj]
  {10.1088/0004-637X/733/1/37}, \href
  {https://ui.adsabs.harvard.edu/abs/2011ApJ...733...37F} {733, 37}

\bibitem[\protect\citeauthoryear{{F{\'a}rn{\'\i}k} \& {Savy}}{{F{\'a}rn{\'\i}k}
  \& {Savy}}{1998}]{1998SoPh..183..339F}
{F{\'a}rn{\'\i}k} F.,  {Savy} S.~K.,  1998, \mn@doi [\solphys]
  {10.1023/A:1005092927592}, \href
  {https://ui.adsabs.harvard.edu/abs/1998SoPh..183..339F} {183, 339}

\bibitem[\protect\citeauthoryear{{F{\'a}rn{\'{\i}}k}, {Hudson}, {Karlick{\'y}}
  \& {Kosugi}}{{F{\'a}rn{\'{\i}}k} et~al.}{2003}]{2003A&A...399.1159F}
{F{\'a}rn{\'{\i}}k} F.,  {Hudson} H.~S.,  {Karlick{\'y}} M.,   {Kosugi} T.,
  2003, \mn@doi [\aap] {10.1051/0004-6361:20021852}, \href
  {http://adsabs.harvard.edu/abs/2003A%26A...399.1159F} {399, 1159}

\bibitem[\protect\citeauthoryear{{Fletcher} \& {Hudson}}{{Fletcher} \&
  {Hudson}}{2008}]{2008ApJ...675.1645F}
{Fletcher} L.,  {Hudson} H.~S.,  2008, \mn@doi [\apj] {10.1086/527044}, \href
  {http://adsabs.harvard.edu/abs/2008ApJ...675.1645F} {675, 1645}

\bibitem[\protect\citeauthoryear{{Fletcher}, {Hannah}, {Hudson}  \&
  {Innes}}{{Fletcher} et~al.}{2013}]{2013ApJ...771..104F}
{Fletcher} L.,  {Hannah} I.~G.,  {Hudson} H.~S.,   {Innes} D.~E.,  2013,
  \mn@doi [\apj] {10.1088/0004-637X/771/2/104}, \href
  {http://adsabs.harvard.edu/abs/2013ApJ...771..104F} {771, 104}

\bibitem[\protect\citeauthoryear{{Freeland} \& {Handy}}{{Freeland} \&
  {Handy}}{1998}]{1998SoPh..182..497F}
{Freeland} S.~L.,  {Handy} B.~N.,  1998, \mn@doi [\solphys]
  {10.1023/A:1005038224881}, \href
  {https://ui.adsabs.harvard.edu/abs/1998SoPh..182..497F} {182, 497}

\bibitem[\protect\citeauthoryear{{Garcia}}{{Garcia}}{1994}]{1994SoPh..154..275G}
{Garcia} H.~A.,  1994, \mn@doi [\solphys] {10.1007/BF00681100}, \href
  {http://adsabs.harvard.edu/abs/1994SoPh..154..275G} {154, 275}

\bibitem[\protect\citeauthoryear{{Graham}, {Hannah}, {Fletcher}  \&
  {Milligan}}{{Graham} et~al.}{2013}]{2013ApJ...767...83G}
{Graham} D.~R.,  {Hannah} I.~G.,  {Fletcher} L.,   {Milligan} R.~O.,  2013,
  \mn@doi [\apj] {10.1088/0004-637X/767/1/83}, \href
  {http://adsabs.harvard.edu/abs/2013ApJ...767...83G} {767, 83}

\bibitem[\protect\citeauthoryear{{Grefenstette} et~al.,}{{Grefenstette}
  et~al.}{2016}]{2016ApJ...826...20G}
{Grefenstette} B.~W.,  et~al., 2016, \mn@doi [\apj]
  {10.3847/0004-637X/826/1/20}, \href
  {http://adsabs.harvard.edu/abs/2016ApJ...826...20G} {826, 20}

\bibitem[\protect\citeauthoryear{{Harra}, {Matthews}, {Culhane}, {Cheung},
  {Kontar}  \& {Hara}}{{Harra} et~al.}{2013}]{2013ApJ...774..122H}
{Harra} L.~K.,  {Matthews} S.,  {Culhane} J.~L.,  {Cheung} M.~C.~M.,  {Kontar}
  E.~P.,   {Hara} H.,  2013, \mn@doi [\apj] {10.1088/0004-637X/774/2/122},
  \href {http://adsabs.harvard.edu/abs/2013ApJ...774..122H} {774, 122}

\bibitem[\protect\citeauthoryear{{Hudson}}{{Hudson}}{1978}]{1978ApJ...224..235H}
{Hudson} H.~S.,  1978, \mn@doi [\apj] {10.1086/156370}, \href
  {https://ui.adsabs.harvard.edu/abs/1978ApJ...224..235H} {224, 235}

\bibitem[\protect\citeauthoryear{{Hudson}, {Strong}, {Dennis}, {Zarro}, {Inda},
  {Kosugi}  \& {Sakao}}{{Hudson} et~al.}{1994}]{1994ApJ...422L..25H}
{Hudson} H.~S.,  {Strong} K.~T.,  {Dennis} B.~R.,  {Zarro} D.,  {Inda} M.,
  {Kosugi} T.,   {Sakao} T.,  1994, \mn@doi [\apjl] {10.1086/187203}, \href
  {http://adsabs.harvard.edu/abs/1994ApJ...422L..25H} {422, L25}

\bibitem[\protect\citeauthoryear{{Hudson}, {Hannah}, {Deluca}  \&
  {Weber}}{{Hudson} et~al.}{2008}]{2008ASPC..397..130H}
{Hudson} H.~S.,  {Hannah} I.~G.,  {Deluca} E.~E.,   {Weber} M.,  2008,
  {Physical Conditions in Coronal Structures About to Flare}.
p.~130

\bibitem[\protect\citeauthoryear{{Jain}, {Joshi}, {Kayasth}, {Dave}  \&
  {Deshpande}}{{Jain} et~al.}{2006}]{2006JApA...27..175J}
{Jain} R.,  {Joshi} V.,  {Kayasth} S.~L.,  {Dave} H.,   {Deshpande} M.~R.,
  2006, \mn@doi [Journal of Astrophysics and Astronomy] {10.1007/BF02702520},
  \href {https://ui.adsabs.harvard.edu/abs/2006JApA...27..175J} {27, 175}

\bibitem[\protect\citeauthoryear{{Kane} \& {Anderson}}{{Kane} \&
  {Anderson}}{1970}]{1970ApJ...162.1003K}
{Kane} S.~R.,  {Anderson} K.~A.,  1970, \mn@doi [\apj] {10.1086/150732}, \href
  {http://adsabs.harvard.edu/abs/1970ApJ...162.1003K} {162, 1003}

\bibitem[\protect\citeauthoryear{{Krucker} et~al.,}{{Krucker}
  et~al.}{2009}]{2009SPIE.7437E..05K}
{Krucker} S.,  et~al., 2009, in \procspie. p. 743705,
  \mn@doi{10.1117/12.827950}

\bibitem[\protect\citeauthoryear{{Lee}, {Petrosian}  \& {McTiernan}}{{Lee}
  et~al.}{1993}]{1993ApJ...412..401L}
{Lee} T.~T.,  {Petrosian} V.,   {McTiernan} J.~M.,  1993, \mn@doi [\apj]
  {10.1086/172929}, \href {http://adsabs.harvard.edu/abs/1993ApJ...412..401L}
  {412, 401}

\bibitem[\protect\citeauthoryear{{Lemen} et~al.,}{{Lemen}
  et~al.}{2012}]{LemenTitleAkin:2012}
{Lemen} J.~R.,  et~al., 2012, \mn@doi [\solphys] {10.1007/s11207-011-9776-8},
  \href {http://adsabs.harvard.edu/abs/2012SoPh..275...17L} {275, 17}

\bibitem[\protect\citeauthoryear{{Lin} et~al.,}{{Lin}
  et~al.}{2002}]{2002SoPh..210....3L}
{Lin} R.~P.,  et~al., 2002, \mn@doi [\solphys] {10.1023/A:1022428818870}, \href
  {http://adsabs.harvard.edu/abs/2002SoPh..210....3L} {210, 3}

\bibitem[\protect\citeauthoryear{{McTiernan}, {Kane}, {Loran}, {Lemen},
  {Acton}, {Hara}, {Tsuneta}  \& {Kosugi}}{{McTiernan}
  et~al.}{1993}]{1993ApJ...416L..91M}
{McTiernan} J.~M.,  {Kane} S.~R.,  {Loran} J.~M.,  {Lemen} J.~R.,  {Acton}
  L.~W.,  {Hara} H.,  {Tsuneta} S.,   {Kosugi} T.,  1993, \mn@doi [\apjl]
  {10.1086/187078}, \href {http://adsabs.harvard.edu/abs/1993ApJ...416L..91M}
  {416, L91+}

\bibitem[\protect\citeauthoryear{{Mrozek} \& {Tomczak}}{{Mrozek} \&
  {Tomczak}}{2004}]{2004A&A...415..377M}
{Mrozek} T.,  {Tomczak} M.,  2004, \mn@doi [\aap] {10.1051/0004-6361:20034598},
  \href {https://ui.adsabs.harvard.edu/abs/2004A&A...415..377M} {415, 377}

\bibitem[\protect\citeauthoryear{{Nagai}}{{Nagai}}{1980}]{1980SoPh...68..351N}
{Nagai} F.,  1980, \mn@doi [\solphys] {10.1007/BF00156874}, \href
  {http://adsabs.harvard.edu/abs/1980SoPh...68..351N} {68, 351}

\bibitem[\protect\citeauthoryear{{Neupert}, {Gates}, {Swartz}  \&
  {Young}}{{Neupert} et~al.}{1967}]{1967ApJ...149L..79N}
{Neupert} W.~M.,  {Gates} W.,  {Swartz} M.,   {Young} R.,  1967, \mn@doi
  [\apjl] {10.1086/180061}, \href
  {https://ui.adsabs.harvard.edu/abs/1967ApJ...149L..79N} {149, L79}

\bibitem[\protect\citeauthoryear{{O'Dwyer}, {Del Zanna}, {Mason}, {Weber}  \&
  {Tripathi}}{{O'Dwyer} et~al.}{2010}]{ODwyerDel-ZannaMason:2010}
{O'Dwyer} B.,  {Del Zanna} G.,  {Mason} H.~E.,  {Weber} M.~A.,   {Tripathi} D.,
   2010, \mn@doi [\aap] {10.1051/0004-6361/201014872}, \href
  {http://adsabs.harvard.edu/abs/2010A%26A...521A..21O} {521, A21}

\bibitem[\protect\citeauthoryear{{Pesnell}, {Thompson}  \&
  {Chamberlin}}{{Pesnell} et~al.}{2012}]{PesnellThompsonChamberlin:2012}
{Pesnell} W.~D.,  {Thompson} B.~J.,   {Chamberlin} P.~C.,  2012, \mn@doi
  [\solphys] {10.1007/s11207-011-9841-3}, \href
  {http://adsabs.harvard.edu/abs/2012SoPh..275....3P} {275, 3}

\bibitem[\protect\citeauthoryear{{Raftery}, {Gallagher}, {Milligan}  \&
  {Klimchuk}}{{Raftery} et~al.}{2009}]{2009A&A...494.1127R}
{Raftery} C.~L.,  {Gallagher} P.~T.,  {Milligan} R.~O.,   {Klimchuk} J.~A.,
  2009, \mn@doi [\aap] {10.1051/0004-6361:200810437}, \href
  {https://ui.adsabs.harvard.edu/abs/2009A&A...494.1127R} {494, 1127}

\bibitem[\protect\citeauthoryear{{Schwartz}, {Csillaghy}, {Tolbert}, {Hurford},
  {McTiernan}  \& {Zarro}}{{Schwartz} et~al.}{2002}]{2002SoPh..210..165S}
{Schwartz} R.~A.,  {Csillaghy} A.,  {Tolbert} A.~K.,  {Hurford} G.~J.,
  {McTiernan} J.,   {Zarro} D.,  2002, \mn@doi [\solphys]
  {10.1023/A:1022444531435}, \href
  {https://ui.adsabs.harvard.edu/abs/2002SoPh..210..165S} {210, 165}

\bibitem[\protect\citeauthoryear{{Serio}, {Reale}, {Jakimiec}, {Sylwester}  \&
  {Sylwester}}{{Serio} et~al.}{1991}]{1991A&A...241..197S}
{Serio} S.,  {Reale} F.,  {Jakimiec} J.,  {Sylwester} B.,   {Sylwester} J.,
  1991, \aap, \href {http://adsabs.harvard.edu/abs/1991A%26A...241..197S} {241,
  197}

\bibitem[\protect\citeauthoryear{{Shih} et~al.,}{{Shih}
  et~al.}{2019}]{2019AGUFMSH33A..08S}
{Shih} A.~Y.,  et~al., 2019, in AGU Fall Meeting Abstracts. pp SH33A--08

\bibitem[\protect\citeauthoryear{{Sim{\~o}es}, {Hudson}  \&
  {Fletcher}}{{Sim{\~o}es} et~al.}{2015a}]{2015SoPh..tmp...50S}
{Sim{\~o}es} P.~J.~A.,  {Hudson} H.~S.,   {Fletcher} L.,  2015a, \mn@doi
  [\solphys] {10.1007/s11207-015-0691-2}, \href
  {http://adsabs.harvard.edu/abs/2015SoPh..290.3625S} {290, 3625}

\bibitem[\protect\citeauthoryear{{Sim{\~o}es}, {Graham}  \&
  {Fletcher}}{{Sim{\~o}es} et~al.}{2015b}]{2015SoPh..290.3573S}
{Sim{\~o}es} P.~J.~A.,  {Graham} D.~R.,   {Fletcher} L.,  2015b, \mn@doi
  [\solphys] {10.1007/s11207-015-0709-9}, \href
  {https://ui.adsabs.harvard.edu/abs/2015SoPh..290.3573S} {290, 3573}

\bibitem[\protect\citeauthoryear{{Sim{\~o}es}, {Reid}, {Milligan}  \&
  {Fletcher}}{{Sim{\~o}es} et~al.}{2019}]{2019ApJ...870..114S}
{Sim{\~o}es} P. J.~A.,  {Reid} H. A.~S.,  {Milligan} R.~O.,   {Fletcher} L.,
  2019, \mn@doi [\apj] {10.3847/1538-4357/aaf28d}, \href
  {https://ui.adsabs.harvard.edu/abs/2019ApJ...870..114S} {870, 114}

\bibitem[\protect\citeauthoryear{{Sterling}, {Hudson}  \&
  {Watanabe}}{{Sterling} et~al.}{1997}]{1997ApJ...479L.149S}
{Sterling} A.~C.,  {Hudson} H.~S.,   {Watanabe} T.,  1997, \mn@doi [\apjl]
  {10.1086/310597}, \href {http://adsabs.harvard.edu/abs/1997ApJ...479L.149S}
  {479, L149}

\bibitem[\protect\citeauthoryear{{Thomas}, {Starr}  \& {Crannell}}{{Thomas}
  et~al.}{1985}]{1985SoPh...95..323T}
{Thomas} R.~J.,  {Starr} R.,   {Crannell} C.~J.,  1985, \mn@doi [\solphys]
  {10.1007/BF00152409}, \href
  {https://ui.adsabs.harvard.edu/abs/1985SoPh...95..323T} {95, 323}

\bibitem[\protect\citeauthoryear{{White}, {Thomas}  \& {Schwartz}}{{White}
  et~al.}{2005}]{2005SoPh..227..231W}
{White} S.~M.,  {Thomas} R.~J.,   {Schwartz} R.~A.,  2005, \mn@doi [\solphys]
  {10.1007/s11207-005-2445-z}, \href
  {http://adsabs.harvard.edu/abs/2005SoPh..227..231W} {227, 231}

\makeatother
\end{thebibliography}

\end{document}